# Oscillations of van Hove singularities spacing induced by sub-Angstrom fluctuations of interlayer spacing in graphene superlattices

Ya-Xin Zhao[1,#], Xiao-Feng Zhou[1,#], Yu Zhang[1,2,3,#], Lin He[1,*]

[1]Center for Advanced Quantum Studies, Department of Physics, Beijing Normal University, Beijing, 100875, People's Republic of China

[2]School of Integrated Circuits and Electronics, MIIT Key Laboratory for Low-Dimensional Quantum Structure and Devices, Beijing Institute of Technology, Beijing 100081, China.

[3]Advanced Research Institute of Multidisciplinary Science, Beijing Institute of Technology, Beijing 100081, China.

[#]These authors contributed equally to this work.

[*]Correspondence and requests for materials should be addressed to Lin He (e-mail: helin@bnu.edu.cn).

**Physical properties of two-dimensional van der Waals (vdWs) structures depend sensitively on both stacking orders and interlayer interactions. Yet, in most cases studied to date, the interlayer interaction is considered to be a "static" property of the vdWs structures. Here we demonstrate that applying a scanning tunneling microscopy (STM) tip pulse on twisted bilayer graphene (TBG) can induce sub-Angstrom fluctuations of the interlayer separation in the TBG, which are equivalent to dynamic vertical external pressure of about 10 GPa on the TBG. The sub-Angstrom fluctuations of the interlayer separation result in large oscillations of the energy separations between two van Hove singularities (VHSs) in the TBG. The period of the oscillations of the VHSs spacing is extremely long, about 500-1000 seconds, attributing to tip-induced local stress in the atomic-thick TBG. Our result provides an efficient method to tune and measure the physical properties of the vdWs structures dynamically.**

Research on exotic electronic properties of van der Waals (vdWs) layered materials has sparked much interest recently [1-5]. In the two-dimensional vdWs systems, the constituent materials, the stacking orders and interlayer interactions are the three dominant factors in determining their physical properties. So far, studies on the physical-property engineering of the vdWs systems with selected constituent materials mainly focus on the relative twist angles between the adjacent layers [6-28] and control of the interlayer twist angle with a precision of about 0.02 ° has already been realized [13]. In the meanwhile, the interlayer interaction of the vdWs materials is usually considered as a fixed value in these studies. Until very recently, several experiments demonstrated that the interlayer spacing of the vdWs materials can be efficiently tuned by hydrostatic pressure due to the weak vdWs forces between adjacent layers [29-33]. These results also demonstrated that the interlayer spacing provides an additional and controllable degree of freedom to tailor the physical properties of the vdWs materials [29-33].

In this Letter, we take advantage of atomic-thick-film nature of the vdWs materials and demonstrate the ability to generate and measure sub-Angstrom fluctuations of the interlayer spacing in twisted bilayer graphene (TBG) by using scanning tunneling microscopy (STM). In the TBG, there are two low-energy van Hove singularities (VHSs), the energy separation of which is extremely sensitive to the interlayer coupling [6-10]. Our experiment indicates that a STM tip pulse on the TBG induces fluctuations of the interlayer spacing, which generates extremely low-frequency oscillations of the VHSs spacing. The extremely sensitive dependence of the VHSs spacing on the interlayer interactions enables us to extract the fluctuations of the interlayer spacing at the sub-Angstrom scale.

Figure 1a shows a schematic of STM measurement on the TBG sample. The period $D$ of the twist-induced moiré pattern is related to the twist angle $\theta$ by $D = a/[2sin(\theta/2)]$, where $a = 0.246\ nm$ is the lattice constant of graphene. In momentum space, the Brillouin zones associated with the two graphene layers are equally rotated by $\theta$, resulting in the Dirac cones centering at different points, $K_1$ and $K_2$, as shown in Fig. 1b. In the absence of interlayer coupling, the low-energy band structure of the TBG consists of the isolated Dirac cones from two graphene layers, intersecting at the energies separated by $\Delta E \approx \hbar v_F \Delta K$, where $\hbar$ is the reduced Planck constant, $v_F$ is the Fermi velocity, $\Delta K = 2|K|sin(\theta/2)$ is the momentum separation of the $K_1$ and $K_2$, and $|K| = 4\pi/3a$. A finite interlayer coupling $t_\theta$ avoids the crossings at the intersections and generates a pair of saddle points in the band structure of the TBG, hence generating two pronounced VHSs in the density-of-states (DOS) [6-11], as shown in Fig. 1b. The energy separations of the two VHSs, $\Delta E_{VHS}$, can be roughly estimated by the continuum model as $\Delta E_{VHS} \approx \hbar v_F \Delta K - 2t_\theta$ [6,7]. Obviously, both the twist angle $\theta$ and the interlayer coupling $t_\theta$ of the TBG play important roles in determining the value of $\Delta E_{VHS}$. According to the *ab initio* results reproduced from the Ref. 10, the value of the $t_\theta$, in the simplest approximation, depends exponentially on the interlayer spacing *d*, as shown in Fig. 1c. Therefore, the interlayer spacing of the TBG can be precisely acquired by measuring the $\Delta E_{VHS}$.

In our experiment, we carried out STM measurements on the TBG at about 77 K. The TBG are obtained by two different methods. One is to transfer monolayer graphene onto the Ni foils that are covered with multilayer graphene [22,27,34,35]. The other is to directly synthesize TBG on Cu-Ni alloys via the chemical vapor deposition (CVD) method [36] (Methods and Figs. S1-S5 [37]). Both systems exhibit quite similar behaviors in the tunable interlayer coupling by using a tip pulse. Because of the atomic-thick-film nature of the graphene systems, recent experiments demonstrated the ability to tune their structures by using local probing tip [27,38-42]. In the bilayer graphene, it is, therefore, possible to change the interlayer separation with the STM tip, as schematically shown in Fig. 2a. Figure 2b-d summarizes representative results of a transfer-assisted TBG on the Ni foil with $\theta$ = 3.02 °. The protuberances and hollows in Fig. 2b of the STM image correspond to the AA and AB/BA stacking regions of the TBG, respectively, with their height differences of about 40 pm. Even though the twist-induced moiré superlattice is observable, the twist-induced low-energy VHSs are absent in the scanning tunneling spectroscopy (STS) spectrum, as shown in Fig. 2d (blue curve), indicating that the interlayer coupling of the TBG is negligible and the topmost graphene sheet behaves as a pristine monolayer graphene. The absence of the low-energy VHSs in the TBG is frequently observed in previous studies [25,43-46], mainly attributing to the enhanced interlayer separations..

In the STS spectra acquired on TBG with $\theta < 5°$ over 100 samples, about 30% exhibits a "V" shape feature and about 70% exhibits two pronounced VHSs, *i.e.*, there is a detectable interlayer coupling in these TBG. For these decoupled TBG, it is interesting to find that we can "switch on" the interlayer coupling by applying a voltage pulse in presence of the feedback loop. As an example, a tip pulse of 4 V for 60-ms duration at the tunneling current of 100 pA on the 3.02 °TBG can dramatically change the electronic properties of the TBG. Firstly, the height differences between the AA and AB/BA stacking regions in the 3.02 °TBG increase to about 90 pm under the same STM measuring conditions, as shown in Fig. 2c. Such a phenomenon arises from the obvious enhancement of the LDOS in the AA regions, which is dominant by the increasing interlayer potential that is associated to the interlayer coupling strength $t_\theta$ [46,47]. The enhanced interlayer coupling is explicitly confirmed in the STS spectrum, as shown in Fig. 2d (red curve). Two low-energy VHSs, arising from the interlayer coupling, emerge in the STS spectrum of the 3.02° TBG. In our experiment, the interlayer interaction of about 70% of the decoupled TBG can be switched on by the tip pulses, indicating that the local interlayer spacing and interlayer coupling of the TBG can be efficiently tuned via a STM tip pulse.

The most surprising result obtained in our experiment is that the energy separations of the two VHSs, $\Delta E_{VHS}$, exhibit large oscillations rather than a fixed constant after the "switching on" the interlayer interaction of the TBG by the tip pulse. Figure 3a shows a representative result of the STS spectra as a function of the time measured in the 3.02° TBG. In each STS spectrum, the voltage is swept from 400 mV to -300 mV with the speed of about 50 mV/s (more parameters for the STS measurements are given in Table S1 [37]). Obviously, the $\Delta E_{VHS}$ exhibits large oscillations and the maximum is almost twice of the minimum. In the TBG, the $\Delta E_{VHS}$ is almost independent of the measured positions [8-10,23-25]. Therefore, the observed large oscillations of the $\Delta E_{VHS}$ should not be arise from any possible drift during the long time STS measurements (Fig. S6 [37]). Figure 3c summarizes the measured $\Delta E_{VHS}$ as a function of the time, which exhibits extremely long period in oscillations. In our experiment, the approximate period in the $\Delta E_{VHS}$ oscillations is about 800 s in the 3.02 °TBG and the STS spectra suddenly become the "V" shape, *i.e.*, the TBG becomes decoupled, after about 3 to 4 periods of the oscillations. Similar result is also obtained in the CVD-grown TBG, as shown in Fig. 3b and 3e. After "switching on" the interlayer coupling of a 3.40 °TBG, large oscillations of the $\Delta E_{VHS}$ with period of about 590 s are observed. Similarly, the STS spectra also change back to the "V" shape suddenly after recording 3 to 4 periods of the oscillations of the VHSs spacing (Fig. S7 [37]). The $\Delta E_{VHS}$ oscillations are quite robust and reproducible in the decoupled TBG (Fig. S8 [37]) and the oscillations could last about 10 periods in some of the TBG when the interlayer coupling is switched on (Fig. S9 [37]). To our best knowledge, such behaviors have never been reported before. According to our STM images of the TBG, we can easily rule out any possible effect of the defects and impurities induced by the tip pulse as the origin of the observed oscillations. The different $\Delta E_{VHS}$ oscillations observed in different TBG samples also help us to remove any possible 'artifacts' originating from the STM tip. To further confirm this, we carried out similar measurements using the same STM tip in the strongly coupled TBG, *i.e.*, the TBG exhibits the two VHSs at the very beginning of STM measurement. In these TBG, the energy separation of the two VHSs, $\Delta E_{VHS}$, is almost independent of the time, as shown in Fig. 3d and 3f, no matter whether a tip pulse is applied or not before the measurement. Therefore, our experiment demonstrated explicitly that a STM tip pulse can efficiently tune the local interlayer coupling and generate oscillations of the VHSs spacing of the TBG.

The above results also indicate that the tip pulse has completely different effects on the decoupled TBG and the strongly coupled TBG. According to previous studies [38,39,45-49], the tip pulse is expected to introduce fluctuations in the flexible atomic-thick membrane through tip-heating effect and weak vdWs forces. The absence of the $\Delta E_{VHS}$ oscillations in the strongly coupled TBG indicates that the two strongly coupled graphene sheets are not moved or they move synchronously after the tip pulse, resulting in the interlayer separation almost unchanged. Whereas, in the decoupled TBG, the topmost graphene sheet may move vertically or both the two decoupled graphene sheets are moved independently after the tip pulse. Therefore, the interlayer separation of the TBG is changed and, consequently, we observe large oscillations of the $\Delta E_{VHS}$. In our analysis, the dependence of the $\Delta E_{VHS}$ on the interlayer coupling strength is obtained according to low-energy band structure of the TBG based on tight-binding model. Figure 4a shows representative low-energy band structures of a 3.15° TBG with different $t_\theta$. Obviously, the energy separations of the two VHSs decrease with increasing $t_\theta$. The extremely sensitive dependence of the $\Delta E_{VHS}$ (the interlayer coupling strength) on the interlayer separation in the TBG (Fig. 1c) enables us to extract the sub-Angstrom fluctuations of the interlayer separation. According to the measured $\Delta E_{VHS}$ oscillations in the 3.02° and 3.40° TBG, the fluctuations of the interlayer spacing in the two TBG can be extracted, as shown in Figs. 4b and 4c. According to our experiment and analysis, the large $\Delta E_{VHS}$ oscillations in the TBG are generated by sub-Angstrom fluctuations of the interlayer separation around the equilibrium interlayer spacing of the bilayer graphene, ~3.2 Angstrom. In our experiment, the observed periods of the $\Delta E_{VHS}$ oscillations range from about 500 s to 1000 s (Fig. 4 and Fig. S9 [37]). The variation of the period may arise from local strain of the TBG, roughness of the substrate, different interactions between the TBG and the substrate, and so on.

The observed extremely low-frequency oscillations of the VHSs spacing can be understood by considering the dynamics of out-of-plane fluctuating graphene sheet subjected to a tip-induced local stress, as reported in Ref. 48. Theoretically, the frequency of graphene vibration can be obtained as $\omega_q = q\sqrt{\frac{\kappa(q^2-q_c^2)}{\rho}}$, where $\rho$ is the mass density of graphene, $\kappa$ is the bending rigidity of graphene, $q_c(A^{-1}) = \sqrt{\frac{|\tau|}{\kappa}}$, and $\tau$ is the stress tensor [48]. When $\tau = 0$ (i.e. the effect of the STM tip is negligible), the frequency reduces to the well-known quadratic dispersion for a vibration graphene membrane, i.e. $\omega_q = \sqrt{\frac{\kappa}{\rho}} q^2$, which is usually in the GHz range [50-52]. For the case that $q$ is near to $q_c$, the frequency can be reduced to extremely low, which results in the observed phenomena in our experiment. In our experiment, a voltage pulse is expected to locally heat the graphene membrane and, therefore, generate a local stress underneath the STM tip. However, a short voltage pulse is impossible to maintain the temperature difference to generate the local stress for tens to hundreds of minutes. It indicates that the continuous tunneling during the STS measurements plays an important role in sustaining the oscillations until the graphene switches back to the weakly coupled regime.

Considering that the long-time low-frequency oscillations of the VHSs spacing, there are three possible mechanisms that can generate continuous strain gradients beneath the STM tip. They are local heating effect, van deer Waals force between the STM tip and the topmost graphene sheet, and the electric field effect induced by the tip during the STS measurements. To further explore this, we carried out STS measurements with different setpoint currents. The oscillation frequency is almost unchanged when the setpoint current is varied from 50 pA to 200 pA, as shown in Fig. S11 [37]. It indicates that the local heating effect and the van der Waals force between the STM tip and the topmost graphene membrane should not be essential to the observed phenomenon because that both of them are strongly dependent on the setpoint current. The most likely mechanism is that the strain gradients in graphene membrane are generated by the tip-induced electric field effect. A local electric field is automatically generated in the region underneath the tip when a bias is applied. According to previous studies, the tip-induced electric field effect can generate a subtle structural distortion of the studied samples [54-57]. With considering that graphene is an atomic-thick ultrasoft film, we believe that such an electric field effect can efficiently introduce a strain gradient during the STS measurements, thus resulting in the low-frequency oscillations of the VHSs spacing. Further works are need to explore how the electric field introduce the local strain in graphene quantitatively.

In summary, we demonstrate the ability to tune and measure the sub-Angstrom fluctuations of the interlayer separation in the TBG with STM tip. Based on recent theoretical result [58], the variations of the interlayer spacing with 10% of the equilibrium interlayer spacing in the TBG are equivalent to a dynamic vertical external pressure on the order of 10 GPa on the TBG. Therefore, the observed tip-induced fluctuations of the interlayer separation allow us to explore exotic electromechanical properties and novel quantum states in vdWs systems in a large dynamic vertical external pressure.

**Acknowledgements**

This work was supported by the National Natural Science Foundation of China (Grant Nos. 11974050, 11674029). L.H. also acknowledges support from the National Program for Support of Top-notch Young Professionals, support from "the Fundamental Research Funds for the Central Universities", and support from "Chang Jiang Scholars Program".

**References**

1. Geim, A. K. & Grigorieva, I. V. Van der Waals heterostructures. *Nature* **499**, 419-425 (2013).

2. Novoselov, K. S., Mishchenko, A., Carvalho, A. & Castro Neto, A. H. 2D materials and van der Waals heterostructures. *Science* **353**, aac9439 (2016).

3. Balents, L., Dean, C. R., Efetov, D. K. & Young, A. F. Superconductivity and strong correlations in moiré flat bands. *Nat. Phys.* **16**, 725-733 (2020).

4. Andrei, E. Y. & MacDonald, A. H. Graphene bilayers with a twist. *Nat. Mater.* **19**, 1265-1275 (2020).

5. Ren, Y., Zhang, Y., Liu, Y. & He, L. Twistronics in graphene-based van der Waals structures. *Chin. Phys. B* **29**, 117303 (2020).

6. Lopes dos Santos, J. M. B., Peres, N. M. R. & Castro Neto, A. H. Graphene Bilayer with a Twist: Electronic Structure. *Phys. Rev. Lett.* **99**, 256802 (2007).

7. Lopes dos Santos, J. M. B., Peres, N. M. R. & Castro Neto, A. H. Continuum model of the twisted graphene bilayer. *Phys. Rev. B* **86,** 155449 (2012).

8. Li, G., Luican, A., Lopes dos Santos, J. M. B., Castro Neto, A. H., Reina, A., Kong, J. & Andrei, E. Y. Observation of Van Hove singularities in twisted graphene layers. *Nat. Phys.* **6**, 109-113 (2010).

9. Yan, W., Liu, M., Dou, R., Meng, L., Feng, L., Chu, Z., Zhang, Y., Liu, Z., Nie, J. & He, L. Angle-Dependent van Hove Singularities in a Slightly Twisted Graphene Bilayer. *Phys. Rev. Lett.* **109**, 126801 (2012).

10. Brihuega, I., Mallet, P., Gonzalez-Herrero, H., Tambly de Laissardiere, G., Ugeda, M. M., Magaud, L., Gomez-Rodriguez, J. M., Yndurain, F. & Veuillen, J. Unraveling the Intrinsic and Robust Nature of van Hove Singularities in Twisted Bilayer Graphene by Scanning Tunneling Microscopy and Theoretical Analysis. *Phys. Rev. Lett.* **109**, 196802 (2012).

11. Cao, Y., Fatemi, V., Demir, A., Fang, S., Tomarken, S. L., Luo, J. Y., Sanchea-Yamagishi, J. D., Watanabe, K., Taniguchi, T., Kaxiras, E., Ashoori, R. C. &

Jarillo-Herrero, P. Correlated insulator behaviour at half-filling in magic-angle graphene superlattices. *Nature* **556**, 80-84 (2018).

12. Cao, Y., Fatemi, V., Fang, S., Watanabe, K., Taniguchi, T., Kaxiras, E. & Jarillo-Herrero, P. Unconventional superconductivity in magic-angle graphene superlattices. *Nature* **556**, 43-50 (2018).

13. Lu, X., Stepanov, P., Yang, W., Xie, M., Aamir, M. A., Das, I., Urgell, C., Watanabe, K., Taniguchi, T., Zhang, G., Bachtold, A., MacDonald, A. H. & Efetov, D. K. Superconductors, orbital magnets and correlated states in magic-angle bilayer graphene. *Nature* **574**, 653-657 (2019).

14. Sharpe, A. L., Fox, E. J., Barnard, A. W., Finney, J., Watanabe, K., Taniguchi, T., Kastner, M. A. & Goldhaber-Gordon, D. Emergent ferromagnetism near three-quarters filling in twisted bilayer graphene. *Science* **365**, 605-608 (2019).

15. Serlin, M., Tschirhart, C. L., Polshyn, H., Zhang, Y., Zhu, J., Watanabe, K., Taniguchi, T., Balents, L. & Young, A. F. Intrinsic quantized anomalous Hall effect in a moiré heterostructure. *Science* **367**, 900-903 (2020).

16. Ponomarenko, L. A., Gorbachev, R. V., Yu, G. L., Elias, D. C., Jalil, R., Patel, A. A., Mishchenko, A., Mayorov, A. S., Mucha-Kruczyski, M., Piot, B. A., Potemski, M., Grigorieva, I. V., Novoselov, K. S., Falko, V. I. & Geim, A. K. Cloning of Dirac fermions in graphene superlattices. *Nature* **497**, 594-597 (2013).

17. Zhang, C., Chuu, C., Ren, X., Li, M., Li, L., Jin, C., Chou, M. & Shih, C. Interlayer couplings, Moiré patterns, and 2D electronic superlattices in $MoS_2/WSe_2$ hetero-bilayers. *Sci. Adv.* **3**, e1601459 (2017).

18. Alexeev, E. M., Ruiz-Tijerina, D. A., Danovich, M., Hamer, M. J., Terry, D. J., Nayak, P. K., Ahn, S., Pak, S., Lee, J., Sohn, J. I., Molas, M. R., Koperski, M., Watanabe, K., Taniguchi, T., Novoselov, K. S., Gorbachev, R. V., Shin, H. S. & Falko, V. I. & Tartakovskii, A. I. Resonantly hybridized excitons in moiré superlattices in van der Waals heterostructures. *Nature* **567**, 81-86 (2019).

19. Bai, Y., Zhou, L., Wang, J., Wu, W., McGilly, L. J., Halbertal, D., BowenLo, C. F., Liu, F., Ardelean, J., Rivera, P., Finney, N. R., Yang, X., Basov, D. N., Yao,

W., Xu, X., Hone, J., Pasupathy, A. N. & Zhu, X. Excitons in strain-induced one-dimensional moiré potentials at transition metal dichalcogenide heterojunctions. *Nat. Mater.* **19**, 1068-1073 (2020).

20. Tran, K., Moody, G., Wu, F., Lu, X., Choi, J., Kim, K., Rai, A., Sanchez, D. A., Quan, J., Singh, A., Embley, J., Zepeda, A., Campbell, M., Autry, T., Taniguchi, T., Watanabe, K., Lu, N., Banerjee, S. K., Silverman, K. L., Kim, S., Tutuc, E., Yang, L., MacDonald, A. H. & Li, X. Evidence for moiré excitons in van der Waals heterostructures. *Nature* **567**, 71-75 (2019).

21. Seyler, K. L., Rivera, P., Yu, H., Wilson, N. P., Ray, E. L., Mandrus, D. G., Yan, J., Yao, W. & Xu, X. Signatures of moiré-trapped valley excitons in $MoSe_2/WSe_2$ heterobilayers. *Nature* **567**, 66-70 (2019).

22. Zhang, Y., Hou, Z., Zhao, Y., Guo, Z., Liu, Y., Li, S., Ren, Y., Sun, Q. & He, L. Correlation-induced valley splitting and orbital magnetism in a strain-induced zero-energy flatband in twisted bilayer graphene near the magic angle. *Phys. Rev. B* **102**, 081403(R) (2020).

23. Li, S., Zhang, Y., Ren, Y., Liu, J., Dai, X. & He, L. Experimental evidence for orbital magnetic moments generated by moiré-scale current loops in twisted bilayer graphene. *Phys. Rev. B* **102**, 121406(R) (2020).

24. Yin, L., Qiao, J., Zuo, W., Li, W. & He, L. Experimental evidence for non-Abelian gauge potentials in twisted graphene bilayers. *Phys. Rev. B* **92**, 081406(R) (2015).

25. Yin, L., Qiao, J., Wang, W., Zuo, W., Yan, W., Xu, R., Dou, R., Nie, J. & He, L. Landau quantization and Fermi velocity renormalization in twisted graphene bilayers. *Phys. Rev. B* **92**, 201408(R) (2015).

26. Yoo, H., Engelke, R., Carr, S., Fang, S., Zhang, K., Cazeaux, P., Sung, S. H., Hovden, R., Tsen, A. W., Taniguchi, T., Watanabe, K., Yi, G., Kim, M., Luskin, M., Tadmor, E. B., Kaxiras, E. & Kim, P. Atomic and electronic reconstruction at the van der Waals interface in twisted bilayer graphene. *Nat. Mater.* **18**, 448-453 (2019).

27. Liu, Y., Su, Y., Zhou, X., Yin, L., Yan, C., Li, S., Yan, W., Han, S., Fu, Z., Zhang, Y., Yang, Q., Ren, Y. & He, L. Tunable Lattice Reconstruction, Triangular Network of Chiral One-Dimensional States, and Bandwidth of Flat Bands in Magic Angle Twisted Bilayer Graphene. *Phys. Rev. Lett.* **125**, 236102 (2020).

28. Arora, H. S., Polski, R., Zhang, Y., Thomosn, A., Choi, Y., Kim, H., Lin, Z., Wilson, I. Z., Xu, X., Chu, J., Watanabe, K., Taniguchi, T., Alicea, J. & Nadj-Perge, S. Superconductivity in metallic twisted bilayer graphene stabilized by $WSe_2$. *Nature* **583**, 379-384 (2020).

29. Yankowitz, M., Watanabe, K., Taniguchi, T., San-Jose, P. & LeRoy, B. J. Pressure-induced commensurate stacking of graphene on boron nitride. *Nat. Commun.* **7**, 13168 (2016).

30. Song, T., Fei, Z., Yankowitz, M., Lin, Z., Jiang, Q., Hwangbo, K., Zhang, Q., Sun, B., Taniguchi, T., Watanabe, K., McGuire, M. A., Graf, D., Cao, T., Chu, J., Cobden, D. H., Dean, C. R., Xiao, D. & Xu, X. Switching 2D magnetic states via pressure tuning of layer stacking. *Nat. Mater.* **18**, 1298-1302 (2019).

31. Li, T., Jiang, S., Sivadas, N., Wang, Z., Xu, Y., Weber, D., Goldberger, J. E., Watanabe, K., Taniguchi, T., Fennie, C. J., Mak, K. F. & Shan, J. Pressure-controlled interlayer magnetism in atomically thin $CrI_3$. *Nat. Mater.* **18**, 1303-1308 (2019).

32. Yankowitz, M., Chen, S., Polshyn, H., Zhang, Y., Watanabe, K., Taniguchi, T., Graf, D., Young, A. F., & Dean, C. R. Tuning superconductivity in twisted bilayer graphene. *Science* **363**, 1059-1064 (2019).

33. Xia, J., Yan, J., Wang, Z., He, Y., Gong, Y., Chen, W., Sum, T. C., Liu, Z., Ajayan, P. M. & Shen, Z. Strong coupling and pressure engineering in $WSe_2$–$MoSe_2$ heterobilayers. *Nat. Phys.* **17**, 92-98 (2021).

34. Suk, J. W., Kitt, A., Magnuson, C. W., Hao, Y., Ahmed, S., An, J., Swan, A. K., Goldberg, B. B. & Ruoff, R. S. Transfer of CVD-grown monolayer graphene onto arbitrary substrates. *ACS Nano* **5**, 6916-6924 (2011).

35. Lin, Y., Jin, C., Lee, J., Jen, S., Suenage, K. & Chin, P. Clean transfer of graphene for isolation and suspension. *ACS Nano* **5**, 2362-2368 (2011).

36. Zhang, Y., Gomez, L., Ishikawa, F. N., Madaria, A., Ryu, K., Wang, C., Badmaev, A. & Zhou, C. Comparison of Graphene Growth on Single-Crystalline and Polycrystalline Ni by Chemical Vapor Deposition. *J. Phys. Chem. Lett.* **1**, 3101-3107 (2010).

37. See supplemental materials for more experimental data, analysis, and further discussion.

38. Zhang, Y., Yang, Q., Ren, Y. & He, L. Observation of phonon peaks and electron-phonon bound states in graphene. *Phys. Rev. B* **100**, 075435 (2019).

39. Li, S., Bai, K., Zuo, W., Liu, Y., Fu, Z., Wang, W., Zhang, Y., Yin, L., Qiao, J. & He, L. Tunneling Spectra of a Quasifreestanding Graphene Monolayer. *Phys. Rev. Appl.* **9**, 054031 (2018).

40. Jia, P., Chen, W., Qiao, J., Zhang, M., Zheng, X., Xue, Z., Liang, R., Tian, C., He, L., Di, Z., Wang, X. Programmable graphene nanobubbles with three-fold symmetric pseudo-magnetic fields. *Nature Commun.* **10**, 3127 (2019).

41. Klimov, N. N., Jung, S., Zhu, S., Li, T., Wright, C. A., Solares, S. D., Newell, D. B., Zhitenev, N. B., Stroscio, J. A. Electromechanical properties of graphene drumheads. *Science* **336**, 1557 (2012).

42. Li, S.-Y., Su, Y., Ren, Y.-N., He, L. Valley polarization and inversion in strained graphene via pseudo-Landau levels, valley splitting of real Landau levels, and confined states. *Phys. Rev. Lett.* **124**, 106802 (2020).

43. Yin, L., Qiao, J., Wang, W., Chu, Z., Zhang, K., Dou, R., Gao, C. L., Jia, J., Nie, J. & He, L. Tuning structures and electronic spectra of graphene layers with tilt grain boundaries. *Phys. Rev. B* **89**, 205410 (2014).

44. Song, Y. J., Otte, A. F., Kuk, Y., Hu, Y., Torrance, D. B., First, P. N., de Heer, W. A., Min, H., Adam, S., Stiles, M. D., MacDonald, A. H. & Stroscio, J. A. High-resolution tunneling spectroscopy of a graphene quartet, *Nature* **467**, 185 (2010).

45. Miller, D. L., Kubista, K. D., Rutter, G. M., Ruan, M., de Heer, W. A., First, P. N. & Stroscio, J. A. Observing the quantization of zero mass carriers in graphene. *Science* **324**, 924 (2009).

46. Yan, W., Liu, M., Dou, R., Meng, L., Feng, L., Chu, Z., Zhang, Y., Liu, Z., Nie, J. & He, L. Angle-dependent van Hove singularities and their breakdown in Twisted Graphene Bilayers. *Phys. Rev. B* **90**, 115402 (2014).

47. Zhang, S., Song, A., Chen, L., Jiang, C., Chen, C., Gao, L., Hou, Y., Liu, L., Ma, T., Wang, H., Feng, X. & Li, Q. Abnormal conductivity in low-angle twisted bilayer graphene. *Sci. Adv.* **6**, eabc5555 (2020).

48. Xu, P., Neek-Amal, M., Barber, S. D., Schoelz, J. K., Ackerman, M. L., Thibado, P. M., Sadeghi, A. & Peeters, F. M. Unusual ultra-low-frequency fluctuations in freestanding graphene. *Nat. Commun.* **5**, 3720 (2014).

49. Flores, F., Echenique, P. M. & Ritchie, R. H. Energy dissipation processes in scanning tunneling microscopy. *Phys. Rev. B* **34**, 2899-2902 (1986).

50. de Andres, P. L., Guinea, F. & Katsnelson, M. I. Density functional theory analysis of flexural modes elastic constants, and corrugations in strained graphene. *Phys. Rev. B* **86**, 225409 (2012).

51. Chen, C. *Graphene NanoElectroMechanical Resonators and Oscillator*. Ph.D. Thesis. Columbia University (2013).

52. Landau, L. D. & Lifshitz, E. M. *Theory of Elasticity* (Pergamon Press, Oxford, 1970).

53. Hansen, O., Ravnkilde, J. T., Quaade, U., Stokbro, K. & Grey, F. Field-Induced Deformation as a Mechanism for Scanning Tunneling Microscopy Based Nanofabrication. *Phys. Rev. Lett.* **81**, 5572-5575 (1998).

54. Klimov, N. N., Jung, S., Zhu, S., Li, T., Wright, C. A., Solares, S. D., Newell, D. B., Zhitenev, N. B. & Stroscio, J. A. Electromechanical Properties of Graphene Drumheads. *Science* **336,** 1557-1561 (2012).

55. Yankowitz, M., Wang, J. I., Birdwell, A. G., Chen, Y., Watanabe, K., Taniguchi, T., Jacquod, P., San-Jose, P., Jarillo-Herrero, P. & LeRoy, B. J. Electric field control of soliton motion and stacking in trilayer graphene. *Nat. Mater.* **13,** 786-789 (2014).

56. Guo, C. X. & Thomson, D. J. Material transfer between metallic tips and surface in the STM. *Ultramicroscopy.* **42-44,** 1452-1458 (1992).

57. Gill, V., Guduru, P. R. & Sheldon, B. W. Electric field induced surface diffusion and micro/nano-scale island growth. *Int. J. Solids Struct.* **45,** 943-958 (2008).

58. Carr, S., Fang, S., Jarllo-Herrero, P. & Kaxiras, E. Pressure dependence of magic twist angle in graphene superlattices. *Phys. Rev. B* **98**, 085144 (2018)

# Figures

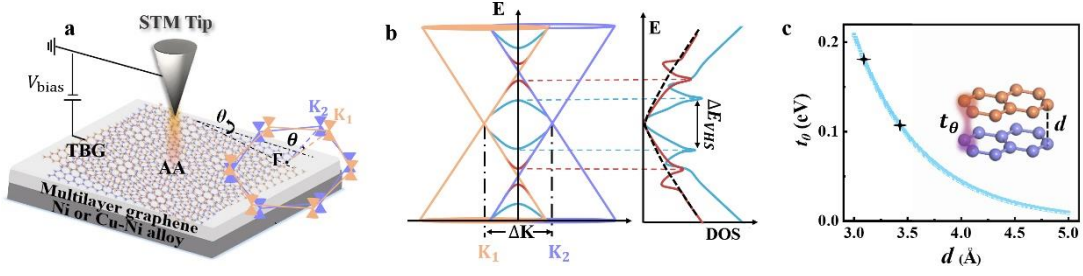

**Fig. 1 |** Schematic atomic structures and low-energy band structures of TBG. **a,** Schematic structure of STM set-up on two misoriented graphene layers with a twist angle $\theta$. Inset: Schematic Brillouin zones associated with the two layers are equally rotated by $\theta$ in momentum space. **b,** Band structures and DOS of TBG. For a fixed twist angle, the energy separations of the two VHSs depend on the interlayer coupling. The low-energy DOS of the TBG with strong, weak, and no interlayer coupling are plotted by the blue, red, and black curves, respectively. **c,** Relation between the interlayer coupling $t_\theta$ and interlayer spacing $d$ of the TBG. The black crosses are extracted from Ref. [10], and the blue line is the fitting result using the equation $t_\theta = Ae^{-B(d-d_{eq})}$, where $A$ and $B$ are the fitting parameters, and $d_{eq} \approx 3.42$ Å is the equilibrium interlayer spacing.

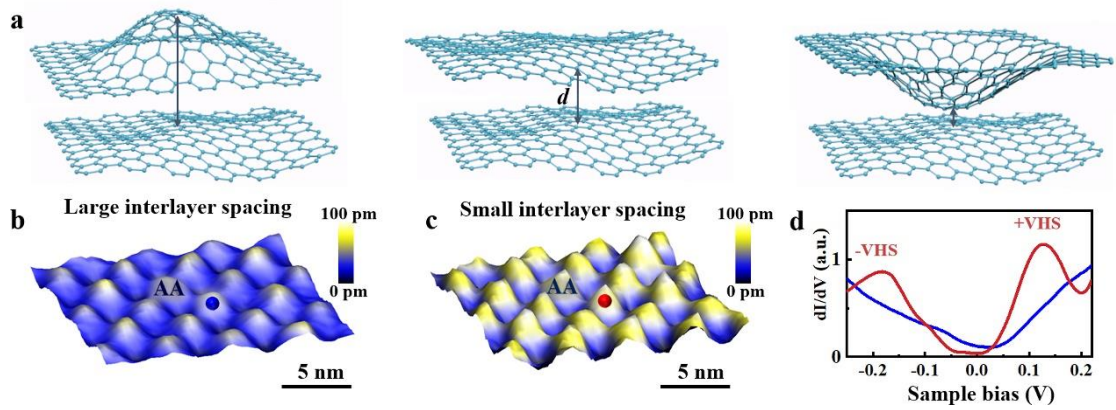

**Fig. 2 | Tunable interlayer spacing and electronic properties of TBG. a,** Schematic structures of TBG with different interlayer separations *d*. **b and c,** Typical STM images of a 3.02° TBG at the same region before and after a STM tip pulse, corresponding to a large and small interlayer spacing, respectively ($V_{bias} = 500\ mV$, $I = 100\ pA$). **d,** Two typical STS spectra of the 3.02° TBG recorded at the AA regions. The blue and red curves are acquired at the dots marked in panels b and c, respectively.

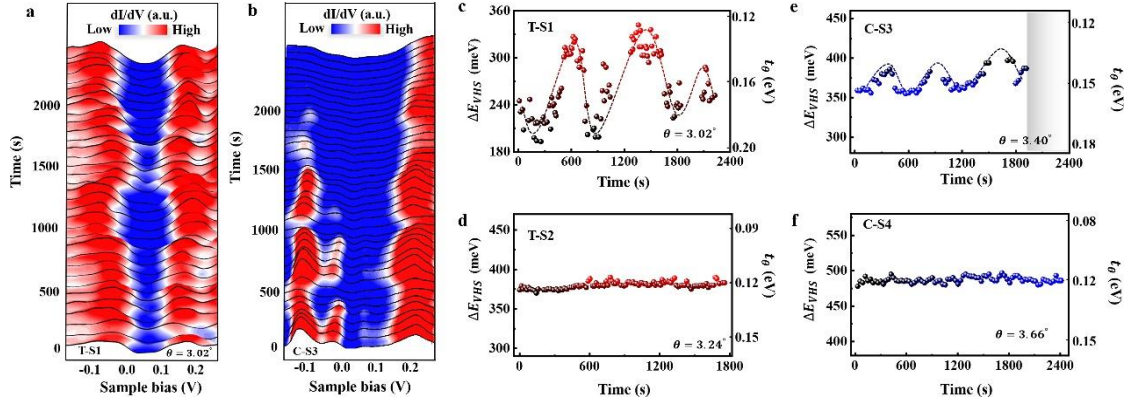

**Fig. 3 | Oscillations of the two VHSs in TBG. a,** Time-dependent STS spectra of the transfer-assisted TBG on Ni foil with $\theta = 3.02°$ (labelled as T-S1). The two VHSs are observable after the STM tip pulses. **b,** Time-dependent STS spectra of the CVD-grown TBG on Cu-Ni alloy with $\theta = 3.40°$ (labelled as C-S3). **c,** The energy separations of the two low-energy VHSs, $\Delta E_{VHS}$, as a function of the measured time for the sample T-S1. The dashed line is guide to eyes. **d,** The measured $\Delta E_{VHS}$ as a function of time for the sample T-S2 on Ni foil. **e,** The measured $\Delta E_{VHS}$ as a function of time for the CVD-grown TBG on Cu-Ni alloy with $\theta = 3.40°$ (labelled as C-S3, the interlayer coupling in this sample is "switched on" by the tip pulses), and the VHSs vanished after ~ 3 to 4 periods (as shown in the grey rectangle). **f,** The measured $\Delta E_{VHS}$ as a function of time for the CVD-grown TBG on Cu-Ni alloy with $\theta = 3.66°$ (labelled as C-S4, the VHSs are observable at the very beginning of the STM characterizations in this sample).

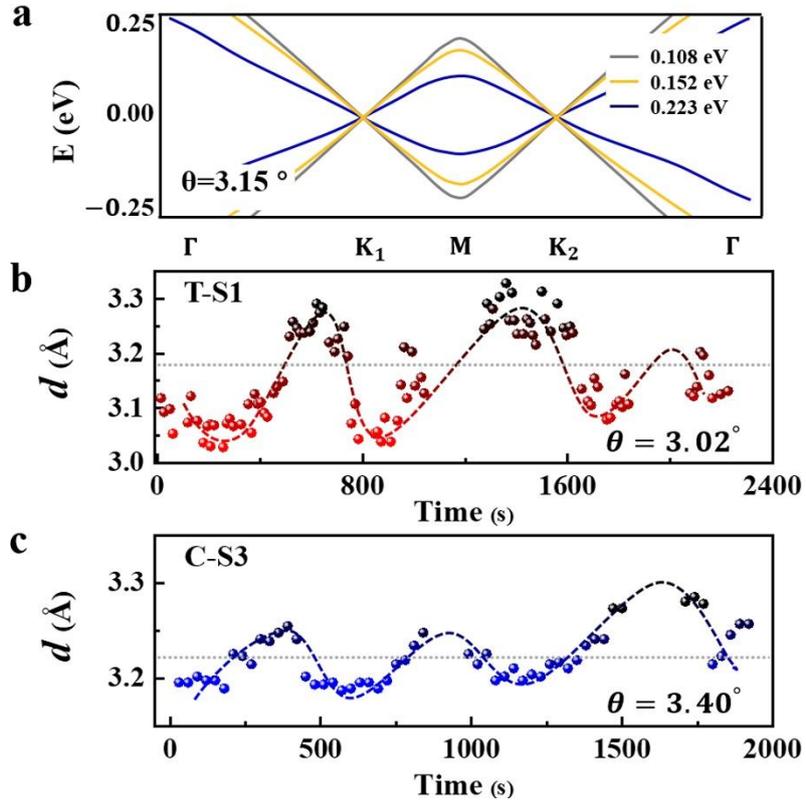

**Fig. 4 | Sub-Angstrom fluctuations of the interlayer spacing in TBG. a,** The low-energy band structures of a 3.15° TBG with different $t_\theta$. **b**, **c** The interlayer separations of TBG as a function of time for the T-S1 and C-S3, respectively. The dashed curves are guide to eyes.